\documentclass[%
 reprint,
 amsmath,amssymb,
 aps,
]{revtex4-2}

\usepackage{graphicx}
\usepackage{dcolumn}
\usepackage{bm}
\usepackage{xcolor}
\usepackage{tabularx}

\begin{document}

\preprint{Physica A}


\title{Signatures of extreme events in the cumulative entropic spectrum}


\author{Ewa A. Drzazga-Szcz{\c{e}}{\'s}niak${^1}$}
\author{Adam Z. Kaczmarek${^2}$}
\author{Marta Kielak${^2}$}
\author{Shivam Gupta$^{3}$}
\author{Jakub T. Gnyp${^4}$}
\author{Katarzyna Pluta${^2}$}
\author{Zygmunt B{\c a}k${^2}$}
\author{Piotr Szczepanik${^5}$}
\author{Dominik Szcz{\c{e}}{\'s}niak${^2}$}
\email{d.szczesniak@ujd.edu.pl}


\affiliation{
${^1}$Department of Physics, Faculty of Production Engineering and Materials Technology, Cz{\c{e}}stochowa University of Technology, 19 Armii Krajowej Ave., 42200 Cz{\c{e}}stochowa, Poland\\
${^2}$Institute of Physics, Faculty of Science and Technology, Jan D{\l}ugosz University in Cz{\c{e}}stochowa, 13/15 Armii Krajowej Ave., 42200 Cz{\c{e}}stochowa, Poland\\
${^3}$EntropyX Labs Pvt. Ltd., Ghaziabad, Uttar Pradesh, 201010, India\\
${^4}$Condensed Matter Spectroscopy Division, Faculty of Mathematics, Physics and Informatics, University of Gda{\'n}sk, Wita Stwosza 57 Str., 80308 Gda{\'n}sk\\
${^5}$Institute of Pricing and Market Analysis, Analitico, 49/8 Królewska Str., 47400 Racib{\'o}rz, Poland
}


\date{\today}


\begin{abstract}

In this study, the cumulative effect of the empirical probability distribution of a random variable is identified as a factor that amplifies the occurrence of extreme events in datasets. To quantify this observation, a corresponding information measure is introduced, drawing upon Shannon entropy for joint probabilities. The proposed approach is validated using selected market data as case studies, encompassing various instances of extreme events. In particular, the results indicate that the introduced cumulative measure exhibits distinctive signatures of such events, even when the data is relatively noisy. These findings highlight the potential of the discussed concept for developing a new class of related indicators or classifiers.

\end{abstract}

\maketitle


\section{Introduction}

The ability to differentiate between volatility inherent to the given data and introduced by a temporal dependency on external variables is necessary for efficient modeling. This is particularly important in the case of significant anomalies, or {\it extreme events}, and their impact on the overall distribution of gathered data. For this purpose, an analytical tool with a solid theoretical foundation and straightforward implementation is required. In search of such a tool it is essential to understand what constitutes an extreme event and what are its characteristics in an {\it ex post} analysis. In loose terms, extreme events are represented by imbalanced data in the time series, which occur irregularly and introduce either extremely low or high values \cite{ding2019}. This implies knowledge of an expected magnitude of the data, with all that lay beyond labeled as \emph{extreme}. One way to express this is by stating that values resulting from extreme events deviate by more than several standard deviations. As such, extreme events span various domains, from science and technology to social studies, and may include sudden outbreaks of devastating infectious diseases, solar flares, extreme weather conditions or financial crises \cite{ramage1980,he2022,weinberg2017}.

This unexpected and complex character of extreme events introduces significant challenges in their theory and modeling. In particular, since all these events often result from strong non-linear interactions across various lengths and time scales, they render conventional perturbative methods less effective \cite{chowdhury2022}. Unfortunately, artificial intelligence has also yet to come to the rescue. Although, there have been attempts to mitigate discussed problems via machine learning (both classical and quantum), not always is there enough data or computational power to perform such simulations \cite{jiang2022, ahmed2024}. Finally, it is important to note that the extreme events contribute to the tails of probabilistic distributions, having minimal effect on mean values but significantly impacting volatility and variance. Interestingly, this opens a promising avenue since one way to analyze volatility is through the measure known as entropy, an analytical concept that also underlines the information theory \cite{shannon1948}. In this sense, entropy estimates uncertainty and randomness of a data allowing to discuss its related fluctuations, distributions, and patterns \cite{dionisio2006,bentes2012,delgado2019,delgado2021,rosser2021,shternshis2022,ormos2014,gupta2024}. As a result, entropy constitutes a potentially highly relevant framework for discussing the impact of sudden events across different fields \cite{sheraz2015,rundle2019,drzazga2023}.

In the present work, it is argued that the entropy can be considered as indicator for extreme events, as motivated by its inherent nature and previous studies \cite{drzazga2023}. Here, this claim is formally justified by following the two point argumentation. First, it is known that the entropy increases as the uncertainty (volatility) of the data rises, meaning that the extreme events should result in the heightened entropy \cite{drzazga2023, gupta2024}. Hence, a simple cumulative process can be considered to further magnify this aspect, leading to a characteristic pattern in the entropic spectrum. This process relates directly to the evolution of the empirical probability distribution. As more data points are taken into account, the observed probability distribution (empirical distribution) gradually converges to the true probability distribution, in accordance with the law of large numbers. Under this assumption the entropy may decrease for a balanced data, signifying reduction in uncertainty and the dominance of a few stable outcomes. By systematically comparing the cumulative distributions generated in this process with the relatively uniform distribution expected for extreme event data, the desired amplification of the latter can be achieved. This approach may be particularly significant as it enables a retrospective analysis, tracing back from the extreme event while cumulatively incorporating data. As a result, it should be possible to develop indicators or techniques for detecting sudden changes in data by considering them as a reference point in time.

This work is organized as follows: Section II introduces the methodology and theoretical background based on the concept of entropy. Section III explores the properties of the data used in this study and provides a detailed analysis of the extreme event signatures within the entropic spectrum. The manuscript concludes in Section IV, which offers a summary and outlines future perspectives. The study is supplemented by an appendix presenting the summary statistics of the analyzed data.

\section{Methodology}

To quantify the discussed cumulative effect we recall the conventional discrete Shannon entropy, given by \cite{shannon1948}:
\begin{equation}
H=-\sum_{i=1}^{n}p(x_{i})\ln(p(x_{i}),
\label{eq01}
\end{equation}
for a total of $n$ outcomes, where:
\begin{equation}
    p(x_{i})=(x_{i+1}-x_i)f(x_{i+1}),
\label{eq02}
\end{equation}
represents the probability of the value $x_i$, in the Riemann approximation, occurring for a discrete random variable of interest. As such, the Eq. (\ref{eq01}) measures information content of data in {\it nats} (meaning the base of logarithm in Eq. (\ref{eq01}) is $e$), accounting for the probability distribution across all possible states. Note that for the discrete case, $n$ constitutes the number of {\it intervals} (known also as {\it bins} or {\it classes}) within the probability distribution. Here, to keep analysis on the same footing, $n$ is assumed after the Vellman formula, which is optimal considering the population and variability of the discussed datasets \cite{dougan2010}. 

Now, lets consider a dataset that contains information about an extreme event, along with preceding data, which is divided into chronologically sorted and equally sized parts. In the extreme limit, it can be qualitatively argued that such blocks exhibit relatively similar probability distribution, except for the portion related to the extreme event. By following in spirit the {\it most biased distribution principle}, familiar in some stochastic processes \cite{gupta2024}, the former dataset blocks will exhibit a bias toward a few stable outcomes, which will be further amplified when their probability distributions are combined. This observation can be quantified by introducing the following {\it cumulative entropy} for the $m$-th dataset part:
\begin{equation}
H_{m}=-\sum_{i=1}^{n}p(x_{i,0}, ... , x_{i,m})\ln{p(x_{i,0}, ... , x_{i,m})},
\label{eq03}
\end{equation}
which is formally a joint entropy for the $m$ blocks of data, where the information is cumulated by adding dataset parts while keeping the same number of outcomes. This means the higher $m$ the more information is contained within the cumulative entropy. This rule applies to all dataset blocks except the one related to an extreme event ($m=0$), for which Eq. (\ref{eq03}) reduces to Eq. (\ref{eq01}). Such process allows to increase the discrepancy in entropy values between higher $m$ terms and reference case when $m=0$. This is to say, the entropy value corresponding to the extreme event can be magnified for its better detection, as initially desired.

Note that, by definition, the $H_{m}$ is non-negative and sub-additive quantity, inheriting these characteristics from Eq. (\ref{eq01}). Moreover, it does not rely on any assumptions about the underlying probability of a data block but instead seeks to uncover its intrinsic characteristics through entropy.

\section{Results and discussion}

To validate the cumulative entropy concept and its underlying rationale, several benchmark datasets are examined. In particular, these consist of market data centered around three key dates, each corresponding to a selected extreme event that occurred in the last decade, {\it i.e.}:

\begin{itemize}

\item June 24th, 2016, marking the announcement of the Brexit referendum results,
\item March 16th, 2020, the global black monday reflecting economic panic due to COVID-19 pandemic,
\item February 24th, 2022, denoting the beginning of Russian invasion on Ukraine.

\end{itemize}

All the above events are captured here in the time series of exchange rates between gold and the U.S. dollar. The total data coverage includes 30 working days before the event, the day of the extreme event itself, and 10 working days after it. Note that the assumed data range directly follows earlier logic of cumulative entropy, which should be determined for the data preceding extreme event, serving as a reference point. This also allows to consider some dates prior this event, to see how corresponding entropic spectrum compares between periods with different reference points. For the same reason the subsequent 10-working-days periods are considered to include the reference points in the future, beyond the extreme event day.

\begin{figure*}[ht!]
\includegraphics[scale=0.8]{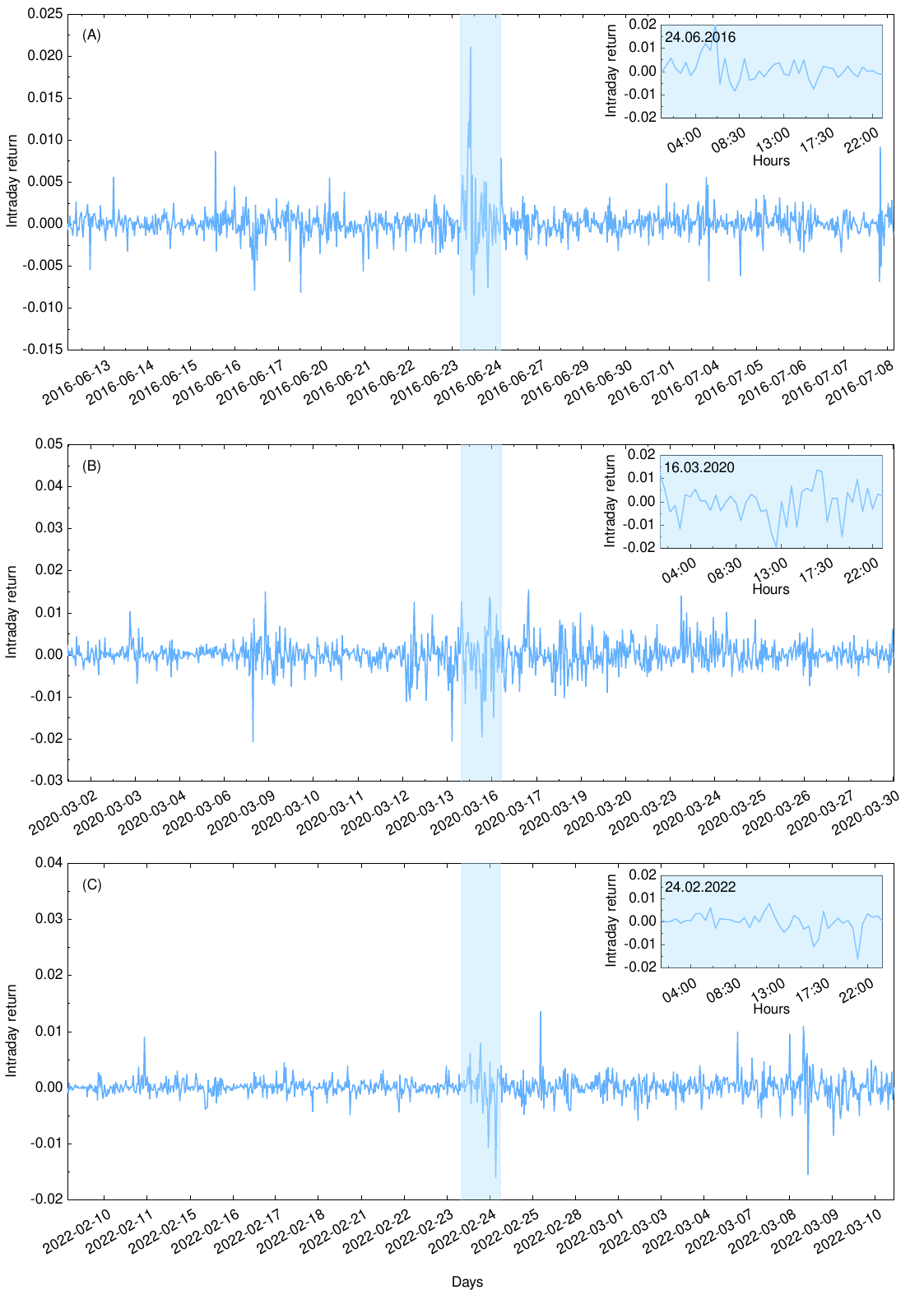}
\caption{The intraday log-returns for the three selected datasets of interest (see text for more details). The data range for each subfigure is restricted to the extreme event day (blue shaded area) $\pm 10$ days. The central regions are additionally magnified in the insets, with the exact dates depicted.}
\label{fig01}
\end{figure*}
\begin{figure*}[ht!]
\includegraphics[scale=0.8]{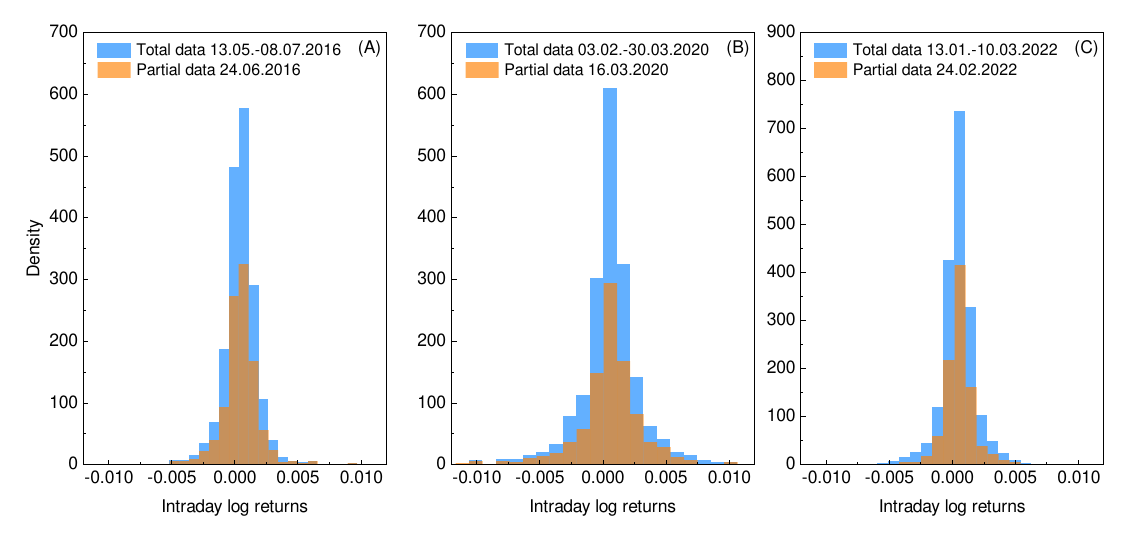}
\caption{The discrete probability distributions for the three selected datasets of interest (see text for more details). Each subfigure depicts function related to the given total dataset spanning range of 41 working days (blue color) and to the partial data for the extreme event day only (orange color).}
\label{fig02}
\end{figure*}

For convenience the information about the time series of interest is encoded via intraday log-returns ($r_{j}$) as:
\begin{equation}
r_j={\rm ln}\frac{p_{j}}{p_{j-1}}\approx\frac{p_{j}-p_{j-1}}{p_{j-1}},
\label{eq04}
\end{equation}
where $p_{j}$ ($p_{j-1}$) is the closing price of an asset on the $j$-th ($j-1$-th) half-hour interval. Such log-returns serve as a stationary time series representation of the price changes, capturing the relative magnitude of intraday fluctuations. 

The graphical representation of the intraday log-returns for the exchange rates between gold and the U.S. dollar is presented in Figs. \ref{fig01} (A)-(C). In a chronological order, they correspond directly to the events listed above. For clarity and transparency, the data range is restricted to the extreme event day $\pm 10$ days. The extreme event day is additionally marked by the blue shaded area and magnified in the inset for further details.

It can be observed that the depicted returns qualitatively exhibit the expected increase in turbulence within the blue shaded area, as evidenced by strong deviations from equilibrium. This effect is particularly pronounced in Figs. \ref{fig01} (A) and (C), which illustrate data behavior for the first and second considered event, respectively. Therein, the transient deviations are nearly four times the equilibrium value. In comparison, the second midterm event is much more noisy across the entire time range, constituting interesting case study when the event of interest is less evident. In what follows, this example extends presented analysis to the cases when events detection is more difficult, allowing to benchmark cumulative entropy concept in complex scenarios. For more details of the considered datasets please refer to Appendix A, where summary statistics are given.

However, in the present paper it is argued that the extreme event is not only reflected in the spectrum of intraday log-returns but can be also observed in the corresponding empirical probability distribution. In Fig. \ref{fig02} (A)-(C), such discrete distributions are presented for each considered total dataset (blue color) and for partial data that refers to the days when the extreme events occur (orange color). By inspecting these results crucial observations can be made that confirm earlier argumentation. In details, the total data is relatively tinner than distribution for the extreme event day. That means information contained within the latter dataset is less ordered and the corresponding outcome is more uncertain. This clearly shows that data for the extreme event day incorporates some randomness into the related total dataset. In other words, as more data is introduced to a dataset, the discrepancy between resulting distribution and the distribution for a single day data, increases. Since this cumulative effect is directly related to the information content within some data, it can be quantified by the entropy.

\begin{figure*}[ht!]
\includegraphics[scale=0.55]{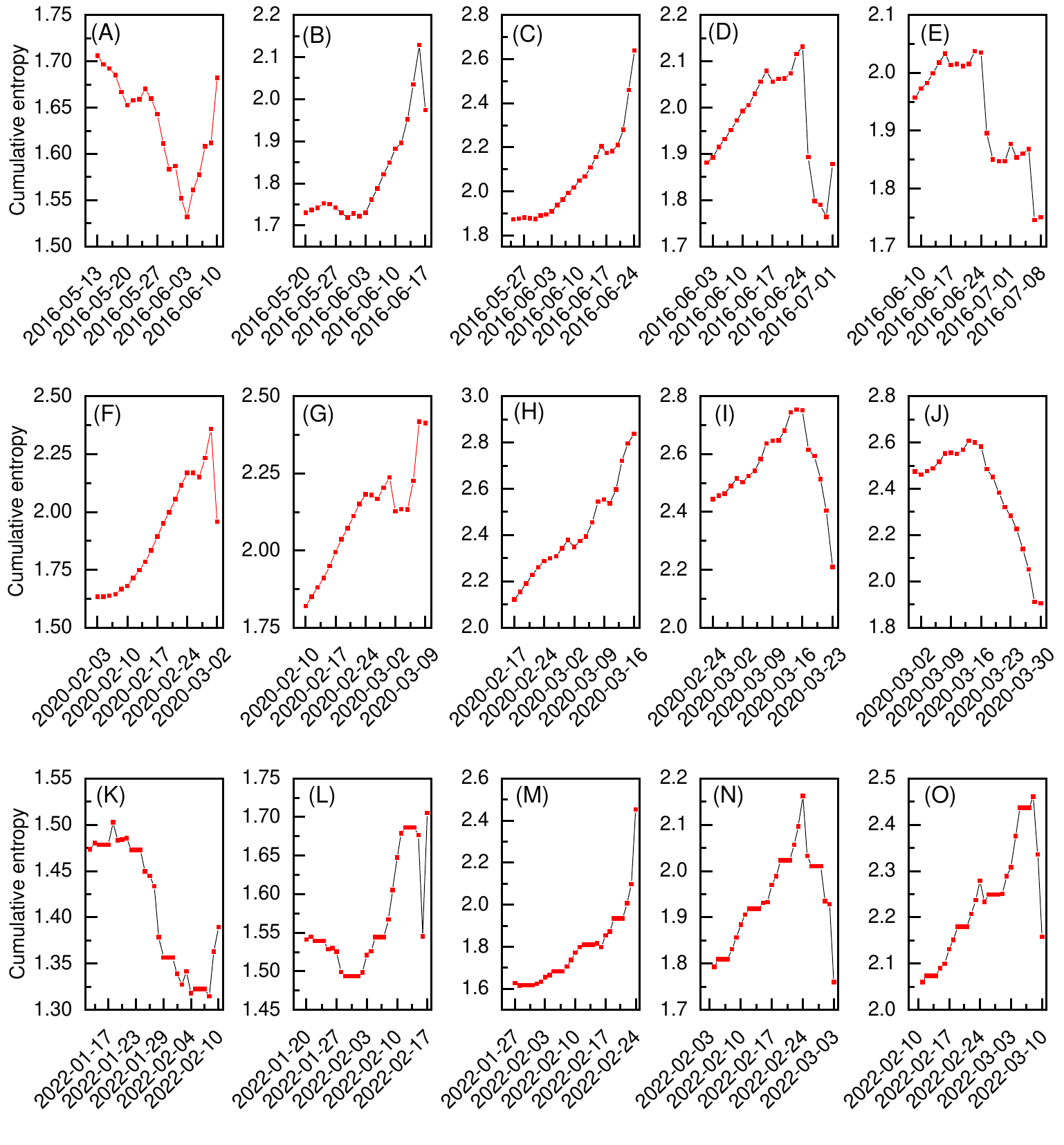}
\caption{The cumulative entropy for the selected datasets of interest (see text for more details). The data range for each subfigure is restricted to 30 working days. The depicted dates refer to the reference point, which is the last data point at each subfigure.}
\label{fig03}
\end{figure*}

In Figs. \ref{fig03} (A)-(O) the behavior of introduced here cumulative entropy for three events of interest is depicted. To capture the underlying features of entropy, the results are presented by employing the 30-day moving window approach. Each window is constructed based on Eq. (\ref{eq03}), assuming the last point as a reference for $m=0$ and every other point to be $m$ days before that event. Note that $m$ numbers the parts, or blocks, that will be compared against each other, as explained in details in the previous section. The analysis begins with interval ending 10 days before extreme event which next sequentially advances every 5 days to conclude with the dataset for the reference point 10 days after the extreme event. In this manner, the middle column (see Figs. \ref{fig03} (C), (H) and (M)) relates to the case when the reference point corresponds to the given extreme event day and the presented results are expected to exhibit some characteristic patterns. Indeed, the results for cumulative entropy, depicted in the middle column, increase steadily as they get closer to the extreme event day, reaching maximum value at reference point. In the first and third row (see Figs. \ref{fig03} (C) and (M)), this increase resembles parabolic behavior and corresponds to the datasets where intraday log-returns at the extreme event day present well indicated deviations from the rest of the data (see Figs. \ref{fig01} (A) and (C)). On the other hand, in the middle row (see Fig. \ref{fig03} (H)) the underlying data is noisy and the behavior of the results obtained for the cumulative entropy is more linear. Still this increase is monotonous without any substantial drops. As such, all three sets of results, given in Figs. \ref{fig03} (C), (H) and (M), clearly present somewhat ordered behavior that differentiate them from the cases when the reference point for calculations is assumed several days after or before the extreme event day. This proves the fact that cumulative entropy may exhibits signatures of interest, even when dataset is relatively noisy.

The observed behavior aligns earlier argumentation, meaning that the entropy value on the day of an extreme event can be magnified by calculating cumulative data for several days prior to this point. In terms of information contained within each of the computed points, the results for $m=0$ clearly correspond to the datasets which are most random and provide most uncertain message. As the results get further from this reference points, more data is added to the corresponding probability distributions and only some of the outcomes dominate. However, what is crucial to note, this behavior is possible only when data with similar probability distributions is cumulated. This is to say, according to the {\it most biased distribution principle}, the deterministic content increases, decreasing related entropy values \cite{gupta2024}. 

In relation to the above, none of other considered time windows follow this process. In other words, it seems like it is difficult to obtain required bias between reference point and the cumulative data when the former does not correspond to the atypical and significant deviation from the equilibrium, familiar for the extreme event. However, with an appropriate time step and careful real-time analysis some early indicators of the upcoming extreme events may be still possible to observe. For example, in market data, investors may make moves that already influence an asset price, which can be reflected in the entropic spectrum.

\section{Summary and conclusions}

In summary, in the present study the analysis was conducted to verify the idea of using entropy measure for detecting extreme events in datasets. It was shown that entropy for joint probabilities can be employed in a systematic manner to magnify data parts or blocks that contain information on an extreme event. In particular, these findings were presented for three datasets of choice, containing market data of exchange rates between gold and the U.S. dollar. Each of the datasets was associated with one extreme event, namely: the announcement of the Brexit referendum results, the global black monday due to COVID-19 pandemic, and the beginning of Russian invasion on Ukraine, respectively. For all three datasets the characteristic signatures in the entropic spectrum were obtained, validating to some extend the proposed theoretical framework.

As a result it can be concluded that the presented method, based on the cumulative entropy, may be beneficial not only for the detection but also classification of extreme events in various datasets. It can serve as a primary or supplementary indicator and classifier that builds upon the underlying distribution of the considered data and the information encapsulated within it. In this manner, the cumulative entropy appears as a universal and comprehensive measure that do not impose any constraints on the corresponding probability distribution, but rather quantifies its underlying and most important features and interdependencies. In this manner, the developed argumentation and obtained results formalize earlier preliminary findings on cumulative entropy concept \cite{drzazga2023}, addressing previously unexplored essential theoretical aspects and providing a corresponding unified framework along with its initial validation.

The above naturally calls for further verification by using large scale data. Of particular interest should be noisy datasets, similar to the one for the global black monday caused by COVID-19 pandemic, where the extraction of information on an extreme event is hindered. Another direction may be implementation of cumulative entropy in the real-time techniques, which deal with the short time windows suggesting opportunity to use developed here measures for early warning systems. The presented study also poses questions on the potential of using the cumulative entropy or the underlying most biased distribution principle in combination with other techniques, similarly to what has been done recently for the geometric Brownian motion process \cite{gupta2024}. One of the promising directions may be be incorporation of the mentioned concepts into the machine learning or deep learning techniques, {\it e.g.} toward improvement of the predictive capabilities of these methods.\\

\noindent {\bf Author Contributions:} Conceptualization, E.A.D.-S. and D.S.; methodology, E.A.D.-S., A.Z.K., M.K. and D.S.; software, E.A.D.-S., A.Z.K., M.K., and D.S.; validation, E.A.D.-S., M.K., S.G, J.T.G., K.P. and P.S.; formal analysis, E.A.D.-S., A.Z.K., M.K., S.G., Z.B. and D.S.; investigation, E.A.D.-S., A.Z.K., M.K., S.G. and D.S.; data curation, E.A.D.-S., M.K., S.G., K.P. and P.S.; writing-original draft preparation, E.A.D.-S., A.Z.K. and J.T.G.; writing-review and editing, E.A.D.-S., A.Z.K., J.T.G. and D.S.; visualization, E.A.D.-S., A.Z.K. and K.P.; supervision, Z.B. and D.S. All authors have read and agreed to the published version of the manuscript.\\

\noindent {\bf Funding:} This research received no external funding.\\

\noindent {\bf Statement:} Not applicable.\\

\noindent {\bf Data Availability Statement:} The original contributions presented in this study are included in the article/supplementary material, while the employed market datasets are publicly available online at www.histdata.com. Further inquiries can be directed to the corresponding author.\\

\noindent {\bf Conflicts of Interest:} The authors declare no conflict of interest.\\

\section{Appendix A. Summary statistics}

The appendix contains supplementary data to the analysis presented in the main text. In particular, in Table \ref{tab01}, the summary statistics of the intraday log-returns are given for three considered datasets that contain information on the analyzed extreme events.

\begin{table*}[]
\renewcommand{\arraystretch}{1.2} 
\setlength{\tabcolsep}{8pt} 
\caption{The summary statistics of the intraday log-returns for the three datasets containing information on extreme events of interest {\it i.e.} the announcement of Brexit referendum results on June 24th 2016, the black Monday on global markets due to the COVID-19 pandemic on March 16th 2020, and the beginning of the Russian invasion on Ukraine on February 24th 2022. The statistics are given for the total period of time considered in the context of each event. Additionally, the partial statistics are presented for five periods analyzed in terms of the cumulative entropy.}
\begin{tabularx}{\textwidth}{|l|X|X|X|X|X|}
\hline
Data & Mean & Minimum & Maximum & Skewness & Kurtosis \\ \hline \hline

\multicolumn{6}{|c|}{Brexit referendum results (24.06.2016)} \\ 
\hline
13.05.-13.06.2016  & 1.560089e-05	& -0.007146 &	0.019459 &	3.102121 &	55.668972 \\
20.05.-20.06.2016  & 2.773924e-05	& -0.008101 &	0.019459 &	2.562111 &	44.722231 \\
25.05-24.06.2016  & 6.775244e-05 & 	-0.008445 &	0.021147 &	3.140278 &	38.333012  \\
02.06.-01.07.2016 & 0.000102 &	-0.008445 &	0.021147 &	3.047951 &	35.344655 \\
09.06.-08.07.2016 & 8.047182e-05	& -0.008445	& 0.021147 &	1.818350 &	22.838294 \\
\hline 
13.05.-08.07.2016 & 4.203620e-05	& -0.008445	& 0.021147 &	2.315603 &	32.974605 \\

\hline \hline

\multicolumn{6}{|c|}{Black monday due to COVID-19 (16.03.2020)} \\ 
\hline
03.02.-03.03.2020 & 3.247892e-05	& -0.014230 & 0.014090 &	-0.551528 &	16.332172 \\
10.02.-09.03.2020 & 7.037389e-05	& -0.020675 & 0.015017 &	-0.856300 &	16.204503 \\
17.02.-16.03.2020 & 6.988914e-05 &	-0.009986 &	0.010257 &	0.001460 &	8.739781 \\ 
24.02.-23.03.2020 & -6.911197e-05	& -0.020675 &	0.015444 &	-0.538275 &	5.644548 \\ 
28.02.-30.03.2020 & -1.000592e-05	& -0.020675 &	0.015444 &	-0.414655 &	5.188489 \\
\hline
03.02.-30.03.2020 
 & 1.431909e-05 &	-0.020675 &	0.015444 &	-0.500483 &	9.635541  \\
 \hline \hline

\multicolumn{6}{|c|}{Russian invasion on Ukraine (24.02.2022)}\\
\hline
13.01.-10.02.2022  & 8.276324e-07	& -0.009664 &	0.006961 &	-0.907719 &	14.53216 \\
20.01.-17.02.2022 & 3.3134e-05	& -0.009664 &	0.009024 &	-0.393835 &	13.431958 \\
27.01.-24.02.2022  & 3.198343e-05	& -0.016004 &	0.009024 &	-1.927835 & 26.631340  \\
03.02.-03.03.2022 & 7.766970e-05 &	-0.016004 &	0.013666 & 	-0.824806 &	20.080578 \\
10.02.-10.03.2022 & 9.916852e-05	& -0.016004 &	0.013666 & 	-0.471262 &	12.495533\\
\hline
13.01.-10.03.2022 & 4.878845e-05 &	-0.016004 &	0.013666 & 	-0.544695 &	18.569120  \\
\hline
\end{tabularx}
\label{tab01}
\end{table*}

\bibliography{apssamp}

\begin{thebibliography}{20}%
\makeatletter
\providecommand \@ifxundefined [1]{%
 \@ifx{#1\undefined}
}%
\providecommand \@ifnum [1]{%
 \ifnum #1\expandafter \@firstoftwo
 \else \expandafter \@secondoftwo
 \fi
}%
\providecommand \@ifx [1]{%
 \ifx #1\expandafter \@firstoftwo
 \else \expandafter \@secondoftwo
 \fi
}%
\providecommand \natexlab [1]{#1}%
\providecommand \enquote  [1]{``#1''}%
\providecommand \bibnamefont  [1]{#1}%
\providecommand \bibfnamefont [1]{#1}%
\providecommand \citenamefont [1]{#1}%
\providecommand \href@noop [0]{\@secondoftwo}%
\providecommand \href [0]{\begingroup \@sanitize@url \@href}%
\providecommand \@href[1]{\@@startlink{#1}\@@href}%
\providecommand \@@href[1]{\endgroup#1\@@endlink}%
\providecommand \@sanitize@url [0]{\catcode `\\12\catcode `\$12\catcode
  `\&12\catcode `\#12\catcode `\^12\catcode `\_12\catcode `\%12\relax}%
\providecommand \@@startlink[1]{}%
\providecommand \@@endlink[0]{}%
\providecommand \url  [0]{\begingroup\@sanitize@url \@url }%
\providecommand \@url [1]{\endgroup\@href {#1}{\urlprefix }}%
\providecommand \urlprefix  [0]{URL }%
\providecommand \Eprint [0]{\href }%
\providecommand \doibase [0]{https://doi.org/}%
\providecommand \selectlanguage [0]{\@gobble}%
\providecommand \bibinfo  [0]{\@secondoftwo}%
\providecommand \bibfield  [0]{\@secondoftwo}%
\providecommand \translation [1]{[#1]}%
\providecommand \BibitemOpen [0]{}%
\providecommand \bibitemStop [0]{}%
\providecommand \bibitemNoStop [0]{.\EOS\space}%
\providecommand \EOS [0]{\spacefactor3000\relax}%
\providecommand \BibitemShut  [1]{\csname bibitem#1\endcsname}%
\let\auto@bib@innerbib\@empty
\bibitem [{\citenamefont {Ding}\ \emph {et~al.}(2019)\citenamefont {Ding},
  \citenamefont {Zhang}, \citenamefont {Pan}, \citenamefont {Yang},\ and\
  \citenamefont {He}}]{ding2019}%
  \BibitemOpen
  \bibfield  {author} {\bibinfo {author} {\bibfnamefont {D.}~\bibnamefont
  {Ding}}, \bibinfo {author} {\bibfnamefont {M.}~\bibnamefont {Zhang}},
  \bibinfo {author} {\bibfnamefont {X.}~\bibnamefont {Pan}}, \bibinfo {author}
  {\bibfnamefont {M.}~\bibnamefont {Yang}},\ and\ \bibinfo {author}
  {\bibfnamefont {X.}~\bibnamefont {He}},\ }\bibfield  {title} {\bibinfo
  {title} {Modeling extreme events in time series prediction},\ }in\ \href
  {https://doi.org/10.1145/3292500.3330896} {\emph {\bibinfo {booktitle}
  {Proceedings of the 25th ACM SIGKDD International Conference on Knowledge
  Discovery \& Data Mining}}},\ \bibinfo {series and number} {KDD '19}\
  (\bibinfo  {publisher} {Association for Computing Machinery},\ \bibinfo
  {year} {2019})\ p.\ \bibinfo {pages} {1114–1122}\BibitemShut {NoStop}%
\bibitem [{\citenamefont {Ramage}(1980)}]{ramage1980}%
  \BibitemOpen
  \bibfield  {author} {\bibinfo {author} {\bibfnamefont {C.}~\bibnamefont
  {Ramage}},\ }\bibfield  {title} {\bibinfo {title} {Sudden events},\
  }\href@noop {} {\bibfield  {journal} {\bibinfo  {journal} {Futures}\ }\textbf
  {\bibinfo {volume} {12}},\ \bibinfo {pages} {268} (\bibinfo {year}
  {1980})}\BibitemShut {NoStop}%
\bibitem [{\citenamefont {He}\ \emph {et~al.}(2022)\citenamefont {He},
  \citenamefont {Wen}, \citenamefont {Huang},\ and\ \citenamefont
  {Ji}}]{he2022}%
  \BibitemOpen
  \bibfield  {author} {\bibinfo {author} {\bibfnamefont {C.}~\bibnamefont
  {He}}, \bibinfo {author} {\bibfnamefont {Z.}~\bibnamefont {Wen}}, \bibinfo
  {author} {\bibfnamefont {K.}~\bibnamefont {Huang}},\ and\ \bibinfo {author}
  {\bibfnamefont {X.}~\bibnamefont {Ji}},\ }\bibfield  {title} {\bibinfo
  {title} {Sudden shock and stock market network structure characteristics: A
  comparison of past crisis events},\ }\href@noop {} {\bibfield  {journal}
  {\bibinfo  {journal} {Technological Forecasting and Social Change}\ }\textbf
  {\bibinfo {volume} {180}},\ \bibinfo {pages} {121732} (\bibinfo {year}
  {2022})}\BibitemShut {NoStop}%
\bibitem [{\citenamefont {Weinberg}\ \emph {et~al.}(2017)\citenamefont
  {Weinberg}, \citenamefont {Andrews},\ and\ \citenamefont
  {Freudenburg}}]{weinberg2017}%
  \BibitemOpen
  \bibfield  {author} {\bibinfo {author} {\bibfnamefont {D.~H.}\ \bibnamefont
  {Weinberg}}, \bibinfo {author} {\bibfnamefont {B.~H.}\ \bibnamefont
  {Andrews}},\ and\ \bibinfo {author} {\bibfnamefont {J.}~\bibnamefont
  {Freudenburg}},\ }\bibfield  {title} {\bibinfo {title} {Equilibrium and
  sudden events in chemical evolution},\ }\href@noop {} {\bibfield  {journal}
  {\bibinfo  {journal} {The Astrophysical Journal}\ }\textbf {\bibinfo {volume}
  {837}},\ \bibinfo {pages} {183} (\bibinfo {year} {2017})}\BibitemShut
  {NoStop}%
\bibitem [{\citenamefont {Chowdhury}\ \emph {et~al.}(2022)\citenamefont
  {Chowdhury}, \citenamefont {Ray}, \citenamefont {Dana},\ and\ \citenamefont
  {Ghosh}}]{chowdhury2022}%
  \BibitemOpen
  \bibfield  {author} {\bibinfo {author} {\bibfnamefont {S.~N.}\ \bibnamefont
  {Chowdhury}}, \bibinfo {author} {\bibfnamefont {A.}~\bibnamefont {Ray}},
  \bibinfo {author} {\bibfnamefont {S.~K.}\ \bibnamefont {Dana}},\ and\
  \bibinfo {author} {\bibfnamefont {D.}~\bibnamefont {Ghosh}},\ }\bibfield
  {title} {\bibinfo {title} {Extreme events in dynamical systems and random
  walkers: A review},\ }\href@noop {} {\bibfield  {journal} {\bibinfo
  {journal} {Physics Reports}\ }\textbf {\bibinfo {volume} {966}},\ \bibinfo
  {pages} {1} (\bibinfo {year} {2022})}\BibitemShut {NoStop}%
\bibitem [{\citenamefont {Jiang}\ \emph {et~al.}(2022)\citenamefont {Jiang},
  \citenamefont {Huang}, \citenamefont {Grebogi},\ and\ \citenamefont
  {Lai}}]{jiang2022}%
  \BibitemOpen
  \bibfield  {author} {\bibinfo {author} {\bibfnamefont {J.}~\bibnamefont
  {Jiang}}, \bibinfo {author} {\bibfnamefont {Z.~G.}\ \bibnamefont {Huang}},
  \bibinfo {author} {\bibfnamefont {C.}~\bibnamefont {Grebogi}},\ and\ \bibinfo
  {author} {\bibfnamefont {Y.~C.}\ \bibnamefont {Lai}},\ }\bibfield  {title}
  {\bibinfo {title} {Predicting extreme events from data using deep machine
  learning: When and where},\ }\href@noop {} {\bibfield  {journal} {\bibinfo
  {journal} {Physical Review Research}\ }\textbf {\bibinfo {volume} {4}},\
  \bibinfo {pages} {023028} (\bibinfo {year} {2022})}\BibitemShut {NoStop}%
\bibitem [{\citenamefont {Ahmed}\ \emph {et~al.}(2024)\citenamefont {Ahmed},
  \citenamefont {Tennie},\ and\ \citenamefont {Magri}}]{ahmed2024}%
  \BibitemOpen
  \bibfield  {author} {\bibinfo {author} {\bibfnamefont {O.}~\bibnamefont
  {Ahmed}}, \bibinfo {author} {\bibfnamefont {F.}~\bibnamefont {Tennie}},\ and\
  \bibinfo {author} {\bibfnamefont {L.}~\bibnamefont {Magri}},\ }\bibfield
  {title} {\bibinfo {title} {Prediction of chaotic dynamics and extreme events:
  A recurrence-free quantum reservoir computing approach},\ }\href@noop {}
  {\bibfield  {journal} {\bibinfo  {journal} {Physical Review Research}\
  }\textbf {\bibinfo {volume} {6}},\ \bibinfo {pages} {043082} (\bibinfo {year}
  {2024})}\BibitemShut {NoStop}%
\bibitem [{\citenamefont {Shannon}(1948)}]{shannon1948}%
  \BibitemOpen
  \bibfield  {author} {\bibinfo {author} {\bibfnamefont {C.~E.}\ \bibnamefont
  {Shannon}},\ }\bibfield  {title} {\bibinfo {title} {A mathematical theory of
  communication},\ }\href@noop {} {\bibfield  {journal} {\bibinfo  {journal}
  {Bell System Technical Journal}\ }\textbf {\bibinfo {volume} {27}},\ \bibinfo
  {pages} {623} (\bibinfo {year} {1948})}\BibitemShut {NoStop}%
\bibitem [{\citenamefont {Dionisio}\ \emph {et~al.}(2006)\citenamefont
  {Dionisio}, \citenamefont {Menezes},\ and\ \citenamefont
  {Mendes}}]{dionisio2006}%
  \BibitemOpen
  \bibfield  {author} {\bibinfo {author} {\bibfnamefont {A.}~\bibnamefont
  {Dionisio}}, \bibinfo {author} {\bibfnamefont {R.}~\bibnamefont {Menezes}},\
  and\ \bibinfo {author} {\bibfnamefont {D.~A.}\ \bibnamefont {Mendes}},\
  }\bibfield  {title} {\bibinfo {title} {An econophysics approach to analyse
  uncertainty in financial markets: an application to the portuguese stock
  market},\ }\href@noop {} {\bibfield  {journal} {\bibinfo  {journal} {The
  European Physical Journal B}\ }\textbf {\bibinfo {volume} {50}},\ \bibinfo
  {pages} {161} (\bibinfo {year} {2006})}\BibitemShut {NoStop}%
\bibitem [{\citenamefont {Bentes}\ and\ \citenamefont
  {Menezes}(2012)}]{bentes2012}%
  \BibitemOpen
  \bibfield  {author} {\bibinfo {author} {\bibfnamefont {S.~R.}\ \bibnamefont
  {Bentes}}\ and\ \bibinfo {author} {\bibfnamefont {R.}~\bibnamefont
  {Menezes}},\ }\bibfield  {title} {\bibinfo {title} {Entropy: A new measure of
  stock market volatility?},\ }in\ \href@noop {} {\emph {\bibinfo {booktitle}
  {Journal of Physics: Conference Series}}},\ Vol.\ \bibinfo {volume} {394}\
  (\bibinfo {organization} {IOP Publishing},\ \bibinfo {year} {2012})\ p.\
  \bibinfo {pages} {012033}\BibitemShut {NoStop}%
\bibitem [{\citenamefont {Delgado-Bonal}(2019)}]{delgado2019}%
  \BibitemOpen
  \bibfield  {author} {\bibinfo {author} {\bibfnamefont {A.}~\bibnamefont
  {Delgado-Bonal}},\ }\bibfield  {title} {\bibinfo {title} {Quantifying the
  randomness of the stock markets},\ }\href@noop {} {\bibfield  {journal}
  {\bibinfo  {journal} {Scientific reports}\ }\textbf {\bibinfo {volume} {9}},\
  \bibinfo {pages} {12761} (\bibinfo {year} {2019})}\BibitemShut {NoStop}%
\bibitem [{\citenamefont {Delgado-Bonal}\ and\ \citenamefont
  {L{\'o}pez}(2021)}]{delgado2021}%
  \BibitemOpen
  \bibfield  {author} {\bibinfo {author} {\bibfnamefont {A.}~\bibnamefont
  {Delgado-Bonal}}\ and\ \bibinfo {author} {\bibfnamefont {{\'A}.~G.}\
  \bibnamefont {L{\'o}pez}},\ }\bibfield  {title} {\bibinfo {title}
  {Quantifying the randomness of the forex market},\ }\href@noop {} {\bibfield
  {journal} {\bibinfo  {journal} {Physica A: Statistical Mechanics and its
  Applications}\ }\textbf {\bibinfo {volume} {569}},\ \bibinfo {pages} {125770}
  (\bibinfo {year} {2021})}\BibitemShut {NoStop}%
\bibitem [{\citenamefont {Rosser~Jr}(2021)}]{rosser2021}%
  \BibitemOpen
  \bibfield  {author} {\bibinfo {author} {\bibfnamefont {J.~B.}\ \bibnamefont
  {Rosser~Jr}},\ }\bibfield  {title} {\bibinfo {title} {Econophysics and the
  entropic foundations of economics},\ }\href@noop {} {\bibfield  {journal}
  {\bibinfo  {journal} {Entropy}\ }\textbf {\bibinfo {volume} {23}},\ \bibinfo
  {pages} {1286} (\bibinfo {year} {2021})}\BibitemShut {NoStop}%
\bibitem [{\citenamefont {Shternshis}\ \emph {et~al.}(2022)\citenamefont
  {Shternshis}, \citenamefont {Mazzarisi},\ and\ \citenamefont
  {Marmi}}]{shternshis2022}%
  \BibitemOpen
  \bibfield  {author} {\bibinfo {author} {\bibfnamefont {A.}~\bibnamefont
  {Shternshis}}, \bibinfo {author} {\bibfnamefont {P.}~\bibnamefont
  {Mazzarisi}},\ and\ \bibinfo {author} {\bibfnamefont {S.}~\bibnamefont
  {Marmi}},\ }\bibfield  {title} {\bibinfo {title} {Measuring market
  efficiency: The {S}hannon entropy of high-frequency financial time series},\
  }\href@noop {} {\bibfield  {journal} {\bibinfo  {journal} {Chaos, Solitons \&
  Fractals}\ }\textbf {\bibinfo {volume} {162}},\ \bibinfo {pages} {112403}
  (\bibinfo {year} {2022})}\BibitemShut {NoStop}%
\bibitem [{\citenamefont {Ormos}\ and\ \citenamefont
  {Zibriczky}(2014)}]{ormos2014}%
  \BibitemOpen
  \bibfield  {author} {\bibinfo {author} {\bibfnamefont {M.}~\bibnamefont
  {Ormos}}\ and\ \bibinfo {author} {\bibfnamefont {D.}~\bibnamefont
  {Zibriczky}},\ }\bibfield  {title} {\bibinfo {title} {Entropy-based financial
  asset pricing},\ }\href@noop {} {\bibfield  {journal} {\bibinfo  {journal}
  {PloS One}\ }\textbf {\bibinfo {volume} {9}},\ \bibinfo {pages} {e115742}
  (\bibinfo {year} {2014})}\BibitemShut {NoStop}%
\bibitem [{\citenamefont {Gupta}\ \emph {et~al.}(2024)\citenamefont {Gupta},
  \citenamefont {Drzazga-Szcz{\c e}{\'s}niak}, \citenamefont {Kais},\ and\
  \citenamefont {Szcz{\c e}{\'s}niak}}]{gupta2024}%
  \BibitemOpen
  \bibfield  {author} {\bibinfo {author} {\bibfnamefont {R.}~\bibnamefont
  {Gupta}}, \bibinfo {author} {\bibfnamefont {E.~A.}\ \bibnamefont
  {Drzazga-Szcz{\c e}{\'s}niak}}, \bibinfo {author} {\bibfnamefont
  {S.}~\bibnamefont {Kais}},\ and\ \bibinfo {author} {\bibfnamefont
  {D.}~\bibnamefont {Szcz{\c e}{\'s}niak}},\ }\bibfield  {title} {\bibinfo
  {title} {Entropy corrected geometric {B}rownian motion},\ }\href@noop {}
  {\bibfield  {journal} {\bibinfo  {journal} {Scientific Reports}\ }\textbf
  {\bibinfo {volume} {14}},\ \bibinfo {pages} {28384} (\bibinfo {year}
  {2024})}\BibitemShut {NoStop}%
\bibitem [{\citenamefont {Sheraz}\ \emph {et~al.}(2015)\citenamefont {Sheraz},
  \citenamefont {Dedu},\ and\ \citenamefont {Preda}}]{sheraz2015}%
  \BibitemOpen
  \bibfield  {author} {\bibinfo {author} {\bibfnamefont {M.}~\bibnamefont
  {Sheraz}}, \bibinfo {author} {\bibfnamefont {S.}~\bibnamefont {Dedu}},\ and\
  \bibinfo {author} {\bibfnamefont {V.}~\bibnamefont {Preda}},\ }\bibfield
  {title} {\bibinfo {title} {Entropy measures for assessing volatile markets},\
  }\href@noop {} {\bibfield  {journal} {\bibinfo  {journal} {Procedia Economics
  and Finance}\ }\textbf {\bibinfo {volume} {22}},\ \bibinfo {pages} {655}
  (\bibinfo {year} {2015})}\BibitemShut {NoStop}%
\bibitem [{\citenamefont {Rundle}\ \emph {et~al.}(2019)\citenamefont {Rundle},
  \citenamefont {Giguere}, \citenamefont {Turcotte}, \citenamefont
  {Crutchfield},\ and\ \citenamefont {Donnellan}}]{rundle2019}%
  \BibitemOpen
  \bibfield  {author} {\bibinfo {author} {\bibfnamefont {J.~B.}\ \bibnamefont
  {Rundle}}, \bibinfo {author} {\bibfnamefont {A.}~\bibnamefont {Giguere}},
  \bibinfo {author} {\bibfnamefont {D.~L.}\ \bibnamefont {Turcotte}}, \bibinfo
  {author} {\bibfnamefont {J.~P.}\ \bibnamefont {Crutchfield}},\ and\ \bibinfo
  {author} {\bibfnamefont {A.}~\bibnamefont {Donnellan}},\ }\bibfield  {title}
  {\bibinfo {title} {Global seismic nowcasting with {S}hannon information
  entropy},\ }\href@noop {} {\bibfield  {journal} {\bibinfo  {journal} {Earth
  and Space Science}\ }\textbf {\bibinfo {volume} {6}},\ \bibinfo {pages} {191}
  (\bibinfo {year} {2019})}\BibitemShut {NoStop}%
\bibitem [{\citenamefont {Drzazga-Szcz{\c e}{\'s}niak}\ \emph
  {et~al.}(2023)\citenamefont {Drzazga-Szcz{\c e}{\'s}niak}, \citenamefont
  {Szczepanik}, \citenamefont {Kaczmarek},\ and\ \citenamefont {Szcz{\c
  e}{\'s}niak}}]{drzazga2023}%
  \BibitemOpen
  \bibfield  {author} {\bibinfo {author} {\bibfnamefont {E.~A.}\ \bibnamefont
  {Drzazga-Szcz{\c e}{\'s}niak}}, \bibinfo {author} {\bibfnamefont
  {P.}~\bibnamefont {Szczepanik}}, \bibinfo {author} {\bibfnamefont {A.~Z.}\
  \bibnamefont {Kaczmarek}},\ and\ \bibinfo {author} {\bibfnamefont
  {D.}~\bibnamefont {Szcz{\c e}{\'s}niak}},\ }\bibfield  {title} {\bibinfo
  {title} {Entropy of financial time series due to the shock of war},\
  }\href@noop {} {\bibfield  {journal} {\bibinfo  {journal} {Entropy}\ }\textbf
  {\bibinfo {volume} {25}},\ \bibinfo {pages} {823} (\bibinfo {year}
  {2023})}\BibitemShut {NoStop}%
\bibitem [{\citenamefont {Do{\u{g}}an}\ and\ \citenamefont
  {Do{\u{g}}an}(2010)}]{dougan2010}%
  \BibitemOpen
  \bibfield  {author} {\bibinfo {author} {\bibfnamefont {N.}~\bibnamefont
  {Do{\u{g}}an}}\ and\ \bibinfo {author} {\bibfnamefont {{\.I}.}~\bibnamefont
  {Do{\u{g}}an}},\ }\bibfield  {title} {\bibinfo {title} {Determination of the
  number of bins/classes used in histograms and frequency tables: a short
  bibliography},\ }\href@noop {} {\bibfield  {journal} {\bibinfo  {journal}
  {{\.I}statistik Ara{\c{s}}t{\i}rma Dergisi}\ }\textbf {\bibinfo {volume}
  {7}},\ \bibinfo {pages} {77} (\bibinfo {year} {2010})}\BibitemShut {NoStop}%
\end{thebibliography}%

\end{document}